# Signature of Electron-magnon Umklapp Scattering in $L1_0$ FePt probed by Thermoelectric Measurements

Xiaoxian Yan,[1] Chang Huai,[2] Hui Xing,[1,*] James P. Parry,[2] Yusen Yang,[1] Guoxiong Tang,[1] Chao Yao,[1] Guohan Hu,[3] Renat Sabirianov,[4] and Hao Zeng[2,†]

[1]*Key Laboratory of Artificial Structures and Quantum Control, and Shanghai Center for Complex Physics, School of Physics and Astronomy, Shanghai Jiao Tong University, Shanghai 200240, China*

[2]*Department of Physics, University at Buffalo, the State University of New York, Buffalo, New York 14260, USA*

[3]*IBM T. J. Watson Research Center, Yorktown Heights, New York 10598*

[4]*Department of Physics, University of Nebraska-Omaha, Omaha, Nebraska 68182*

We report unconventional thermoelectric power (Seebeck coefficient, $S$) in $L1_0$ structured FePt films. The temperature dependence of $S$ can be well fitted by a phenomenological expression consisting of electron diffusion and magnon-drag contributions. Interestingly, the magnon drag coefficient carries an opposite sign to that of electron diffusion, revealing a dominant contribution from the elusive electron-magnon Umklapp scattering. DFT calculations identify several bands crossing the Brillouin zone boundaries, facilitating the Umklapp process. The large spin-orbit coupling in FePt results in strong mixing of majority and minority spins among some of those bands, greatly enhancing the electron-magnon scattering.

* huixing@sjtu.edu.cn

† haozeng@buffalo.edu

The transfer of spin angular momentum from conduction electrons to magnetization, and its reverse effect, namely magnetization dynamics induced spin and charge current, can be harnessed for applications in information processing and storage [1,2], such as spin transfer torque magnetic random access memory (STT-MRAM) [3], current driven domain wall motion for racetrack memory [4], and thermally or microwave driven spin currents for logic devices [5]. The emerging fields of spintronics, spin-caloritronics and magnonics call for a deeper understanding of the interplay between spin dynamics and transport properties in magnetic materials [6-9]. It has been shown theoretically that the parameterized spin transfer torque can be determined by the magnon-drag thermopower [10]. However, experimental reports of magnon-drag thermopower are rare [11-13] as compared to phonon-drag thermopower. The rarity of magnon dominated thermoelectric transport behavior also hinders theoretical development. Oftentimes theoretical interpretations of the electron-magnon interactions did not take into account the underlying electronic structure of the materials studied [14].

Observation of magnon-drag effect in transport measurements remains elusive due in part to the fact that contributions from electron diffusion and phonon drag often mask the contribution from magnons. Because of this, ingenious design of thermopile structures was developed to elucidate the effects of magnons, by suppressing the electron and phonon contributions in thermopower [15]. Because of the potential for improving thermoelectric performance by exploiting the magnetic degree of freedom [16-19], thermoelectric transport in magnetic materials, including transition metal and alloys, gained renewed attention recently [20-23]. $L1_0$ FePt holds a special place among these materials, with a large magnetocrystalline anisotropy $K_u > 10^7$ J/m$^3$ highly desirable for overcoming the superparamagnetic effect in magnetic data storage and spintronic devices [24-28]. Understanding electron-magnon interactions in FePt may help to expand its applications in spin transfer- based devices such as spin-orbit torque MRAM [29-31]. As demonstrated by Mihai *et al.* [32], the electron-magnon scattering contributes to the magnetoresistance (MR) of FePt, leading to a linear dependence of MR on external field. While similar behavior has been found in 3*d* metals at high fields [33], its presence in FePt at lower fields is due to its high $K_u$ [34,35]. FePt possesses large magnetic moment and strong spin-orbit coupling (SOC) that mixes spin-up and spin-down electronic states. The combination of both facilitates the spin-flip scattering due to electron-magnon interactions. FePt can thus be an ideal system for exploring the magnon-drag effect in its thermoelectric response.

Here we report the first study of temperature-dependent thermopower behavior of $L1_0$ FePt films. Temperature dependence of the Seebeck coefficient $S$ at low temperatures reveals prominent magnon-drag component. Surprisingly, both the electron diffusion and magnon drag exhibit unconventional behaviors. The electron diffusion thermopower ($S_d$) is found to be positive as opposed to being negative in conventional metals with nearly free electrons. The concave (*i.e.* hole-like) curvature of corresponding Fermi surfaces in FePt contributes to the positive $S$. More interestingly, the magnon-drag thermopower ($S_m$) shows an opposite sign to that of $S_d$. This is attributed to the electron-magnon Umklapp scattering (U process), which although theoretically predicted, has not been experimentally observed [36,37]. Such unusual thermopower behavior of FePt distinguishes itself from other known ferromagnetic systems such as Fe, Co and Ni [11,36], and roots in the specific electronic structure of the material.

FePt polycrystalline films were deposited by DC magnetron sputtering of [Fe 5Å|Pt 5Å]$_{10}$ multilayer films on (001) MgO substrates. The films were subsequently annealed for 1 hour at 550 °C in $H_2/N_2$ atmosphere to form the 10 nm thick $L1_0$ ordered FePt alloy. For all electron transport and thermoelectric measurements, the magnetic fields were applied perpendicular to the film plane. In-plane sample resistance was measured using the *ac* lock-in module in a Quantum Design Physical Property Measurement System (PPMS). Thermopower measurements were performed in the PPMS at temperatures between 2 and 300 K using steady-state technique. The temperature gradient, maintained at around 0.3 K/mm, was applied in-plane and monitored by a pair of type-E differential thermocouple. Band structure calculations were performed using DFT-based method implemented in the plane-wave density functional code VASP. Projector augmented wave PAW pseudopotentials were used for Fe and Pt. The generalized gradient approximation of Perdew-Burke-Ernzerhof form is used for the exchange-correlation functional. We used a 16×16×16 *k*-points sampling for 2 atoms super cell. We set the plane-wave cut-off energy to 350 eV and choose the convergence criteria for energy of $10^{-6}$ eV. SOC is included in the calculation.

The temperature dependent sheet resistance $R_\square$ of the 10 nm $L1_0$ FePt film is shown in Fig. 1(a), which exhibits typical metallic behavior. A notable feature is the $T^2$ behavior of the resistance at low temperatures up to 80 K, as shown in the insert of Fig. 1(b). This $T^2$ dependence in ferromagnetic metals has been documented as a signature of electron-magnon scattering decades ago [37]. The anomalous Hall effect (AHE) and magnetoresistance of FePt shown in Fig. 1(b, c) corroborate the existence of electro-magnon scattering. The AHE reveals a magnetic hysteresis loop with a nearly unity magnetic remanence, which excludes complications from anisotropic magnetoresistance and domain wall propagation. In this case the linear dependence of resistance on the magnetic field shown in Fig. 1(e) originates from the electron-magnon scattering [32], as mentioned before. The electron transport firmly establishes the important role of electron-magnon scattering in FePt, hinting its possible detection by thermoelectric measurements.

A representative Seebeck coefficient $S(T)$ for the FePt films at zero magnetic field is shown in Fig. 2. $S$ is found to be negative at higher temperatures and decreases in magnitude upon cooling. Interestingly, $S$ changes its sign to positive at $T_{\mathrm{cross}} = 44$ K. A peak in $S(T)$ developes at around 11 K, which subsequently decreases towards zero as $T$ approaches zero, as required by the third law of thermodynamics. In this work, we focus on the thermopower behavior below 100 K. The higher-$T$ thermopower involves complex scattering processes, which is beyond the scope of our paper.

At relatively low temperatures, the Seebeck coefficient $S$ of a metal consists of several contributions, namely electron diffusion $S_{\mathrm{d}}$, electron-phonon scattering (phonon-drag) $S_{\mathrm{ph}}$, and in magnetic materials electron-magnon scattering (magnon-drag) $S_{\mathrm{m}}$. $S_{\mathrm{d}}$ scales linearly with $T$, and carries a sign mainly determined by its band curvature. The phonon contribution typically shows a temperature dependence proportional to the phonon entropy $T^3$. Likewise, the magnon-drag part essentially resembles the phonon-drag physics, with its magnitude scaling with the magnon entropy of $T^{3/2}$ due to the magnon dispersion relation different from that of the phonon.

The presence of a peak at $\sim 10$ K implies that either phonon- or magnon-drag might exist. To elucidate the dominating contributions in thermopower, we first attempted to fit the experimental data at $T < T_{\mathrm{cross}}$ using the empirical expression: $S = S_{\mathrm{d}} + S_{\mathrm{ph}} = a \cdot T + b \cdot T^3$. The result was not satisfactory. Moreover, the $S(T)$ peak magnitude decreases with increasing magnetic field, as seen in Fig. 2(b). This is opposite to the expected field dependence of a phonon peak [38], but consistent with the magnon-drag behavior. Applying a magnetic field will suppress the magnon population, thus suppress the magnon peak [32,33]. We therefore attribute the observed peak to magnon-drag effect. The temperature dependence of $S$ can thus be described using the expression $S = S_{\mathrm{d}} + S_{\mathrm{m}} = \alpha \cdot T + \beta \cdot T^{3/2}$, where $\alpha$ is the electron diffusion coefficient and $\beta$ the magnon-drag coefficient. The data are replotted as $S/T$ *vs*. $T^{1/2}$ in Fig. 3. It can be seen that the data at zero and moderate magnetic fields (up to 6 T) are well described by a linear behavior for $T < T_{\mathrm{cross}}$, confirming a major contribution from magnon-drag in the low-$T$ thermopower, in addition to the electron diffusion part that exists at all temperatures. Apparent deviation appears at higher fields and lower temperatures (> 6 T and < 10 K) in Fig. 3 (b). Both low temperature and high field strongly suppress the magnon population, thus the thermoelectric transport is dominated by electron diffusion, which leads to nearly constant $S/T$. Meanwhile, we see that the coefficient $\alpha$ also decreases with increasing field, which is a direct consequence of the curved electron trajectory in the presence of a perpendicular magnetic field. On the other hand, the absence of a phonon-drag peak is reasonable considering the existence of substantial chemical disorder at our annealing temperature [39].

Extracting the coefficients $\alpha$ and $\beta$, as shown in Fig. 3, results in two important findings. (i) The electron diffusion coefficient $\alpha$ is positive; (ii) The magnon drag coefficient $\beta$ is negative, which is opposite to the sign of $\alpha$. Both are counterintuitive and not typically observed in normal metals or ferromagnets. In the rest of the paper, we will focus on these two observations.

Electron diffusion Seebeck coefficient can be expressed [12,40] as $S_{\mathrm{d}} = -\frac{\pi^2 k_{\mathrm{B}}^2 T}{3|e|}\left(\frac{1}{A}\frac{\partial A}{\partial E} + \frac{1}{l}\frac{\partial l}{\partial E}\right)_{E=E_{\mathrm{F}}}$, as a result of the Boltzmann equation, where $A$ is the Fermi surface area and $l$ is the electron mean free path. It is clear that Fermi surface topology and the carrier mean free path together determine the sign of $S_{\mathrm{d}}$. The scattering term $\frac{1}{l}\frac{\partial l}{\partial E}$ is usually positive since electrons with higher energy are harder to be scattered (The scattering term is trickier in noble metal, which can become negative, see *e.g.*, [41,42]). For nearly free electron systems, the parabolic bands show positive $\frac{1}{A}\frac{\partial A}{\partial E}$, therefore $S_{\mathrm{d}}$ is negative. However, $S_{\mathrm{d}}$ can change sign if the Fermi surfaces are strongly distorted, leading to negative $\frac{1}{A}\frac{\partial A}{\partial E}$, as observed in noble metals [43-45].

To understand the positive sign of the electron diffusion term, we need to examine the detailed band structure. The term $\frac{1}{A}\frac{\partial A}{\partial E}$ in FePt is determined for each band with DFT calculations. The band structure and Fermi surface of $L1_0$ structured FePt are shown in Fig. 4. The seven bands crossing the Fermi level are labeled as bands 1 to 7. Bands 1 and 4 have mainly Fe $3d$-character, bands 6 and 7 have Pt $5d$-character and are strongly spin mixed. The rest of the bands are hybridized states of Fe and Pt. As can be seen in Fig. 4, for all the bands except band 2, the Fermi surface area increases with decreasing energy (Fig. 4(c)), while decreases with increasing energy (Fig. 4(d)). Band 2 shows opposite trend. The total $\frac{1}{A}\frac{\partial A}{\partial E}$ of all the bands is found to be −0.52 $eV^{-1}$. The estimated contribution from the band structure to the diffusive thermopower is $-\frac{\pi^2 k_B^2}{3|e|}\left(\frac{1}{A}\frac{\partial A}{\partial E}\right)_{E=E_F}$ ~0.013 $\mu V/K^2$, which constitutes a small and positive contribution to the diffusion coefficient $\alpha$. The rest contribution to a positive $\alpha$ should come from the scattering term $\frac{1}{l}\frac{\partial l}{\partial E}$, similar to those found in noble metals [41,42].

In the presence of positive diffusive thermopower $S_d$, the negative magnon drag $S_m$ remains nontrivial. Recent theoretical work shows that both hydrodynamic contribution concerning the Gilbert damping $\alpha'$ and a Berry phase contribution concerning the adiabatic damping $\beta'$ are involved in the magnon drag [46,47]. The sign of the magnon drag $S_m$ depends on $\beta'/\alpha'$ whose value is beyond direct estimation. So far, interpretation of transition-metal ferromagnets using this model has not yielded physically reasonable damping parameters [21]. It is also worth to emphasize that the linear relation between $S/T$ and $T^{1/2}$ at different magnetic fields all hold up to $T = T_{cross}$ as shown in Fig. 3, indicating its sensitivity to scattering details, rather than magnetic damping. In light of these observations, we propose that the opposite signs between those of $S_d$ and $S_m$ can be understood by the electron-magnon Umklapp scattering process. The electron-magnon scattering process can be described by $k - k' - q = g$, where $k$ and $k'$ are the momentum vectors of an electron before and after the scattering, $q$ is the momentum of the magnon and $g$ is the reciprocal lattice vector. $g = 0$ denotes the normal scattering process (N process); while $g \neq 0$ is the Umklapp process (U process). The N process gives rise to a drag thermopower with the same sign as the electron diffusion thermopower, since the change of the electron vector $\Delta k$ is parallel to $q$. While for the U process, as depicted in Fig. 5, the result is different. An electron with vector $k$ is scattered by a magnon and absorbs the magnon vector $q$; it then reaches the neighboring zone in the repeated Brillouin zone with vector $k'$. The reciprocal lattice vector $g$ transfers $k'$ back to the first Brillouin zone, resulting in a change of the electron momentum vector $\Delta k$ that is opposite to the magnon vector $q$. This process therefore leads to a drag thermopower $S_m$ with the opposite sign to that of electron diffusion $S$.

To the best of our knowledge, the electron-magnon Umklapp scattering has not been experimentally observed. The question arises then is what makes FePt special? According to the discussions above, there are two criteria for observing the elusive electron-magnon Umklapp scattering: (i) Electron-magnon scattering should be strong, *i.e.* there should be spin-up and spin-down electronic states accessible by magnon scattering and such scattering should play a dominant role over phonon scattering; and (ii) Similar to the phonon U processes, there should be bands crossing the Brillouin zone boundaries to permit electron-magnon U process at low temperatures. Therefore, the mechanism of the observed U process should be rooted in the detailed band structure of FePt. As can be seen from Fig. 5 (a), band 2 cannot be responsible for the U process at low temperatures. This is because band 2 does not cross the zone boundary, and therefore a minimum magnon vector $q_{min}$ is required for the U process to occur, essentially freezing out this process at low temperatures. Band 1 cannot contribute to the U process for the same reason. On the other hand, bands 3-7 would allow for the electron-magnon Umklapp scattering to happen, since these bands are connected through the zone boundaries with no required $q_{min}$ for the U scattering process. The connected bands across zone boundaries and the Umklapp scattering process are shown in Fig. 5(b) and 5(c) for bands 3 and 4, respectively. Among the bands where Umklapp scattering is allowed, only bands 3 and 4 are important since they show avoided band crossings due to strong SOC. This leads to hybridization between the majority and minority spin states, thus enhancing the electron-magnon scattering [48]. The enhanced electron-magnon Umklapp scattering competes with electron diffusion, thus leads to the emergence of a peak in thermopower. The extracted magnon-drag thermopower coefficient $\beta$ is about 0.1 $\mu V/K^{5/2}$, which is an order of magnitude higher than that of Fe (0.016 $\mu V/K^{5/2}$). The strong electron-magnon interactions suggest that FePt could be a potential candidate for spin-charge-heat interconversion through spin angular momentum transfer.

In summary, we report the unusual thermopower behavior in $L1_0$ FePt films. Magnon-drag thermopower is found to be a major contribution to the total thermopower. The sign of magnon-drag thermopower is found to be opposite to that of electron diffusion thermopower. We identified several bands crossing the zone boundaries, with a mixing of the majority and minority spins, as a consequence of the strong SOC in FePt. These bands enhance the electron- magnon Umklapp scattering, leading to opposite signs between the two terms. It can be expected that electron-magnon Umklapp scattering may be observable in other 3*d*-4*d* and 3*d*-5*d* ferromagnetic alloys with strong spin orbit coupling, such as CoPt and FePd.

The work at SJTU was supported by National Key Projects for Research & Development of China (Grant No. 2019YFA0308602), National Natural Science Foundation of China (Grant No. 11804220), and Natural Science Foundation of Shanghai (Grant No. 20ZR1428900). H.X. also acknowledges additional support from a Shanghai talent program. The work at SUNY-Buffalo was supported by the NSF (Grant No. ECCS-2042085). Computations resources were provided by the University of Nebraska Holland Computing Center.

DATA AVAILABILITY

The data that support the findings of this study are available from the corresponding author upon reasonable request.

Figure 1:

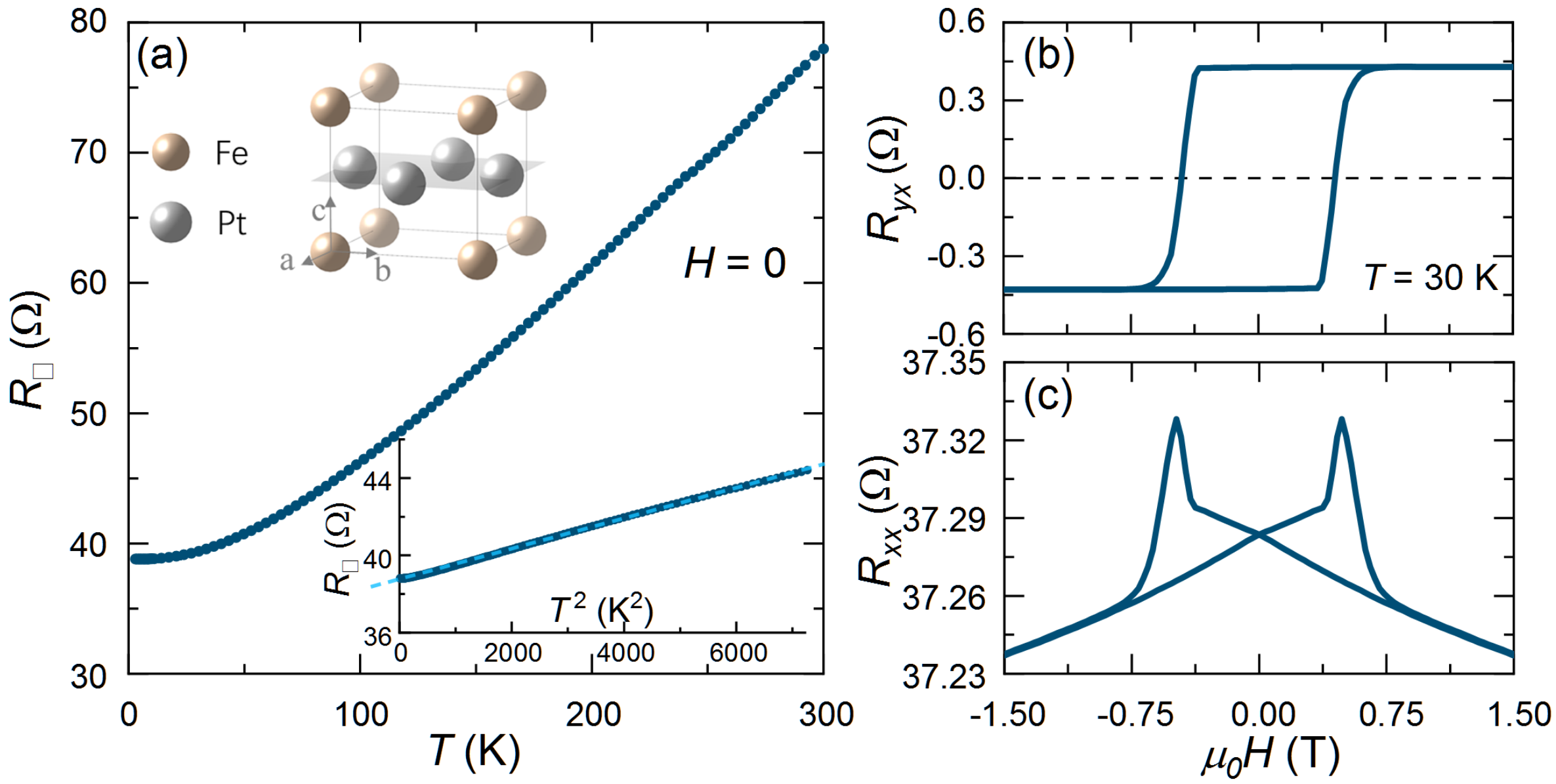

Fig. 1: (color online) (a) (a) The temperature dependence of zero-field sheet resistance $R_\square$ of the 10 nm FePt grown on MgO substrates. Top inset shows the lattice structure of $L1_0$ FePt. Lower inset shows the sheet resistance scales linearly with $T^2$ up to 80 K. Blue dash lines are linear fitting. (b, c) The Anomalous Hall resistance and the magnetoresistance of the 10 nm FePt film at 50 K. The magnetic field is applied perpendicular to the film plane.

Figure 2:

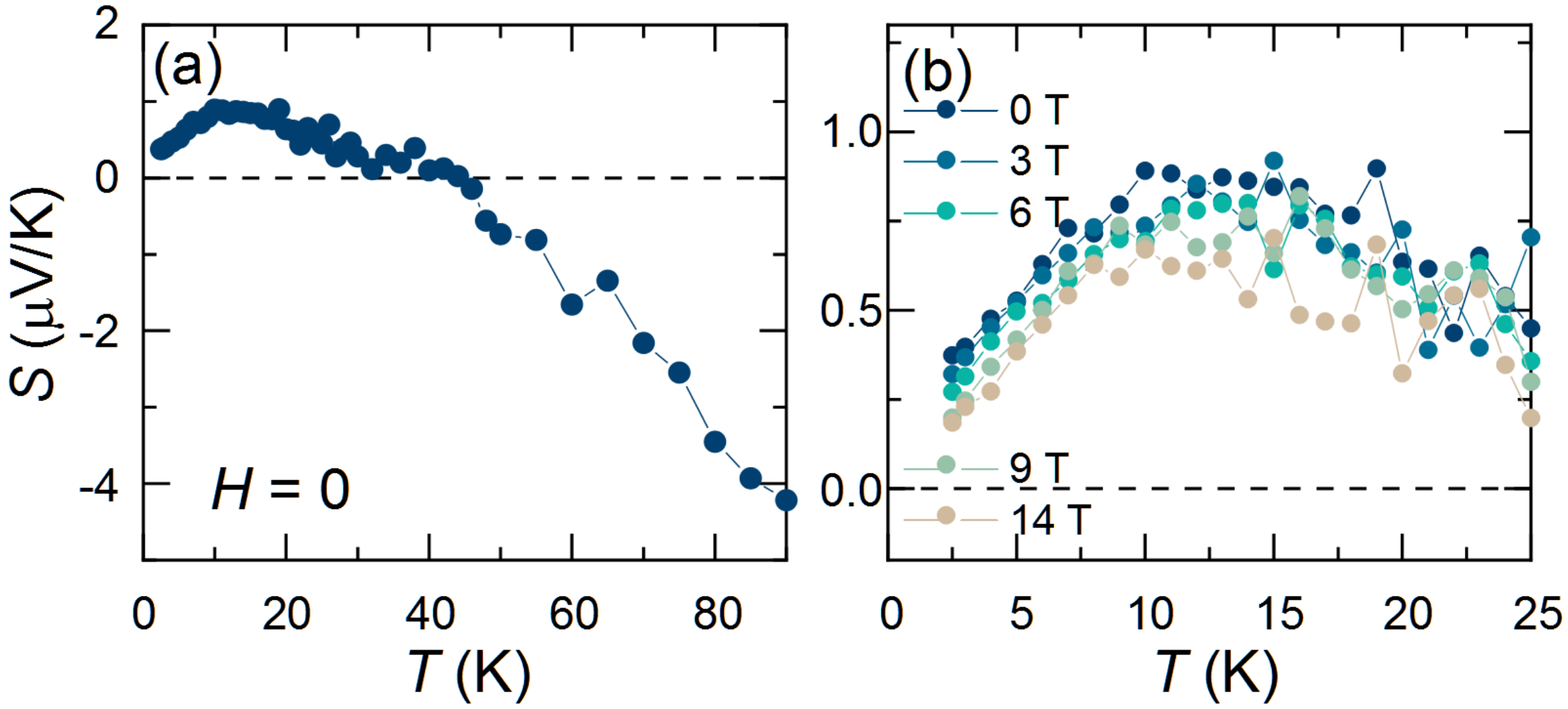

FIG. 2: (color online) (a) The temperature dependence of zero-field thermopower of the 10 nm FePt film on MgO substrates. (b, c) The low-temperature thermopower of the 10 nm FePt film at different magnetic fields. The magnetic field is applied perpendicular to the film plane.

Figure 3:

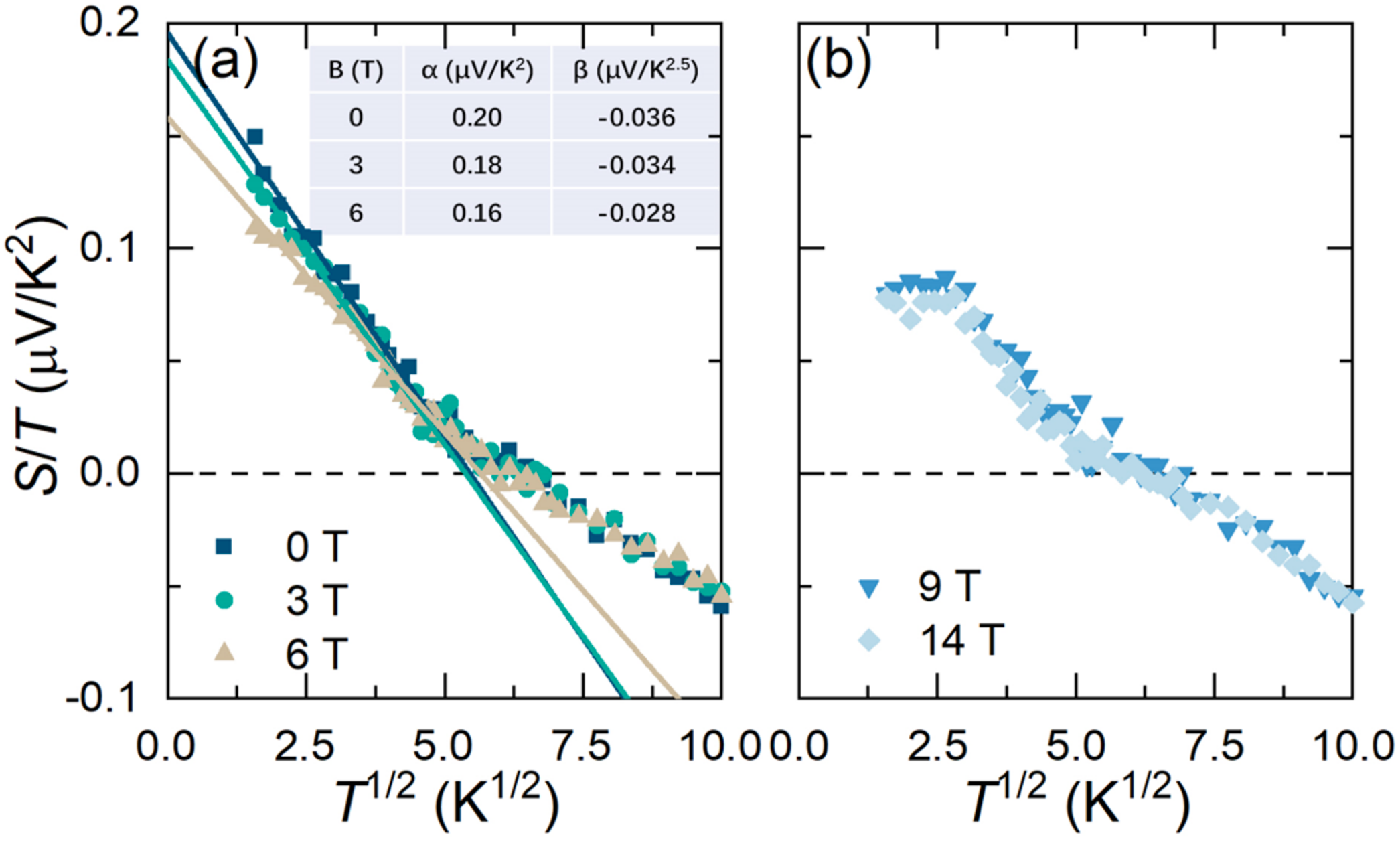

FIG. 3: (color online) (a) The low-temperature thermopower of the 10 nm FePt film grown on MgO substrates at different magnetic fields. The magnetic field is applied perpendicular to the film plane. The solid lines are fitting using $S/T = \alpha + \beta \cdot T^{1/2}$. The extracted fitting parameters $\alpha$ and $\beta$ are shown in the insets.

Figure 4:

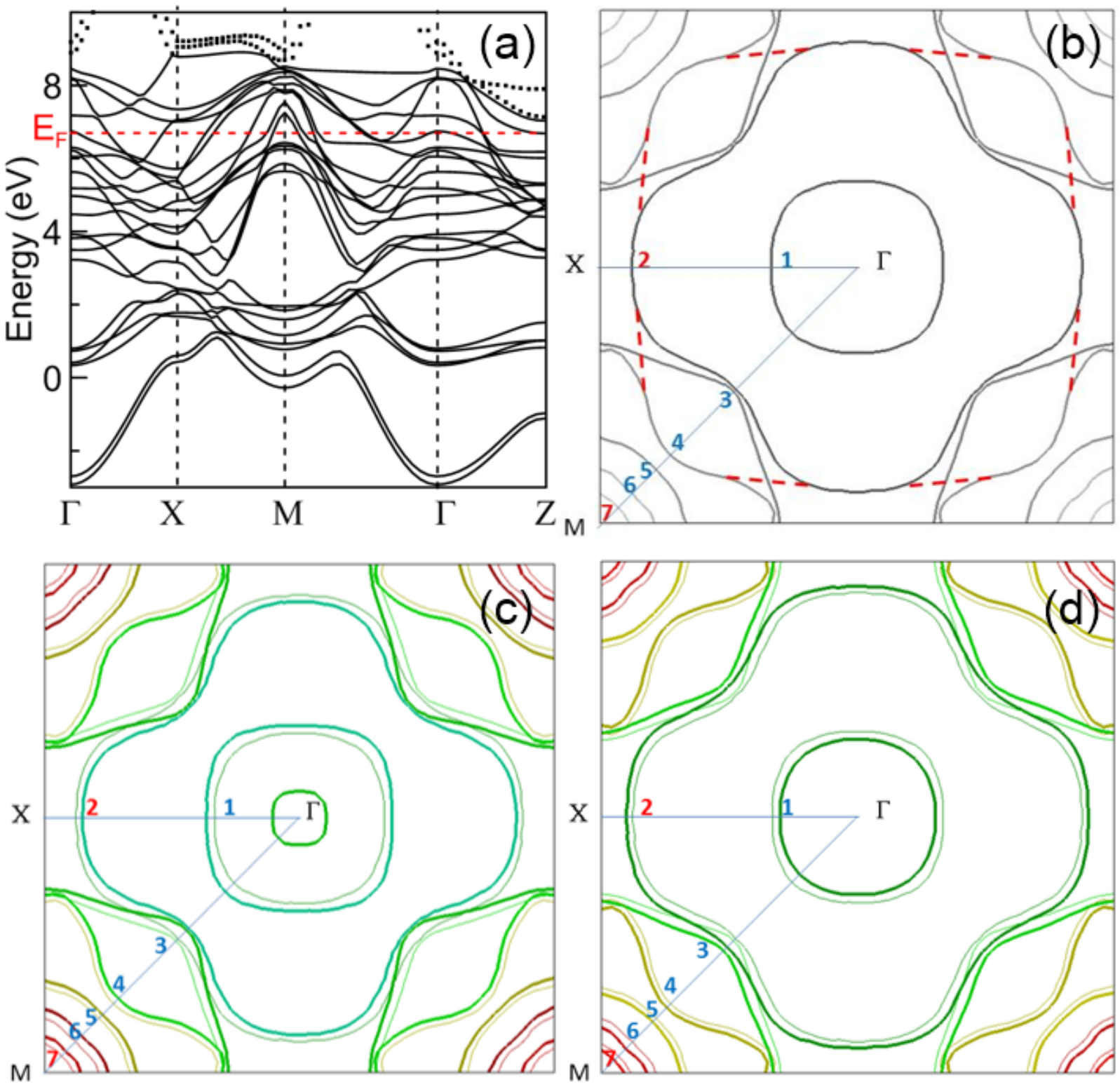

FIG. 3: (color online) (a) Band structure of FePt with spin-orbit coupling included. There are seven bands crossing the Fermi level. Notice that the Fermi surface pockets near M point reduces in size with an effective increase of Fermi energy. (b) Cross-section of Fermi surface of FePt. There are seven sheets crossing *xy*-plane numbered in order of remoteness from Γ point. Red/blue colored numbers indicate spin majority/minority character in relation to calculation without SOC. Sheets 2, 3 and 4 have mixed majority-minority states. Red dashed line represents schematically the spin majority Fermi surface sheet in calculations without SOC. (c,d) Contour plot of Fermi surface sheets in $z = 0$ plane (thin lines) and two isovalued lines (thick lines) plotted for $E_{\mathrm{F}} - 0.1$ eV (c) and $E_{\mathrm{F}} + 0.1$ eV (d). This figure shows that $\frac{1}{A}\frac{\partial A}{\partial E}$ is negative because the increase in energy causes most of Fermi surface sheets to reduce their area A. Total $\frac{1}{A}\frac{\partial A}{\partial E}$ summing up all the bands is $-0.52$ $\mathrm{eV}^{-1}$.

Figure 5:

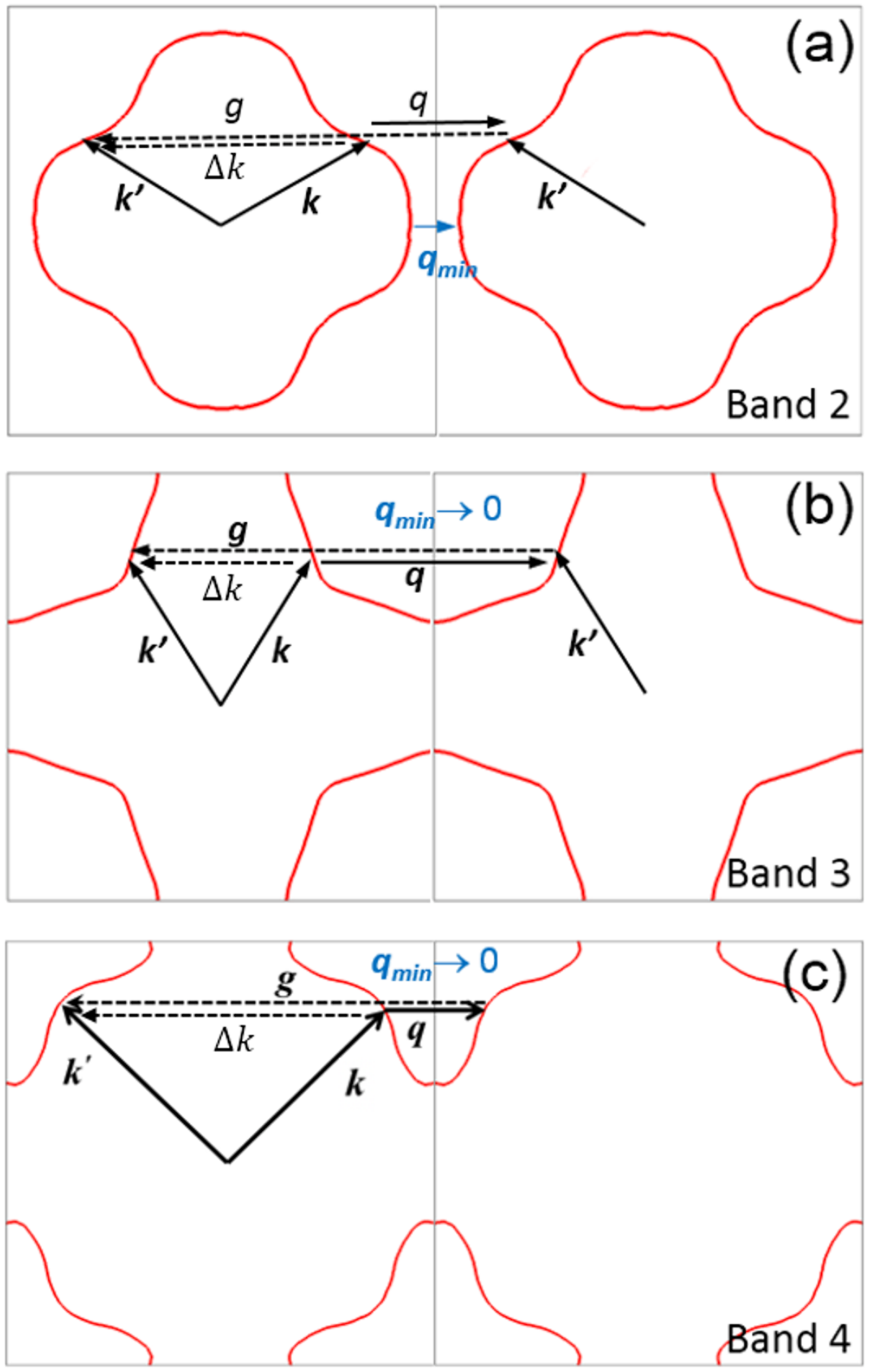

FIG. 5: (color online) Electron-magnon Umklapp scattering process in Band 2, 3 and 4. Bands 2 requires a finite $q_{\text{min}}$ for U process, while bands 3 and 4 permit U process at low temperatures. $q$, $g$ and $\Delta k$ denote the momentum of the magnon, the reciprocal lattice vector and the change of the electron vector, respectively.